\documentclass[%
 aip,
 amsmath,amssymb,
 reprint,%
]{revtex4-1}

\usepackage{graphicx}
\usepackage{dcolumn}
\usepackage{bm}

\usepackage[utf8]{inputenc}
\usepackage[T1]{fontenc}
\usepackage{mathptmx}
\usepackage{etoolbox}

\makeatletter
\def\@email#1#2{%
 \endgroup
 \patchcmd{\titleblock@produce}
  {\frontmatter@RRAPformat}
  {\frontmatter@RRAPformat{\produce@RRAP{*#1\href{mailto:#2}{#2}}}\frontmatter@RRAPformat}
  {}{}
}%
\makeatother

\preprint{AIP/123-QED}

\begin{document}

\title{Classification of complex local environments in systems of particle shapes through shape-symmetry encoded data augmentation}

\author{Shih-Kuang (Alex) Lee}
\affiliation{Department of Material Science and Engineering, University of Michigan, Ann Arbor, MI 48109, USA.}

\author{Sun-Ting Tsai}
\affiliation{Department of Chemical Engineering, University of Michigan, Ann Arbor, MI 48109, USA.}

\author{Sharon C. Glotzer*}
\email{sglotzer@umich.edu}
\affiliation{Department of Chemical Engineering, University of Michigan, Ann Arbor, MI 48109, USA.}
\affiliation{Biointerfaces Institute, University of Michigan, Ann Arbor, MI 48109, USA.}
\date{\today}

\begin{abstract}
\section*{Abstract}
Detecting and analyzing the local environment is crucial for investigating the dynamical processes of crystal nucleation and shape colloidal particle self-assembly. Recent developments in machine learning provide a promising avenue for better order parameters in complex systems that are challenging to study using traditional approaches. However, the application of machine learning to self-assembly on systems of particle shapes is still underexplored. To address this gap, we propose a simple, physics-agnostic, yet powerful approach that involves training a multilayer perceptron (MLP) as a local environment classifier for systems of particle shapes, using input features such as particle distances and orientations. Our MLP classifier is trained in a supervised manner with a shape symmetry-encoded data augmentation technique without the need for any conventional roto-translations invariant symmetry functions. We evaluate the performance of our classifiers on four different scenarios involving self-assembly of cubic structures, 2-dimensional and 3-dimensional patchy particle shape systems, hexagonal bipyramids with varying aspect ratios, and truncated shapes with different degrees of truncation. The proposed training process and data augmentation technique are both straightforward and flexible, enabling easy application of the classifier to other processes involving particle orientations. Our work thus presents a valuable tool for investigating self-assembly processes on systems of particle shapes, with potential applications in structure identification of any particle-based or molecular system where orientations can be defined.
\end{abstract}

\maketitle

\section{Introduction}
\label{sec:introduction}

Self-assembly is studied extensively in such fields as physics, chemistry, materials science, chemical engineering and biology \cite{boles2016self,whitesides2002beyond}. A fundamental process involving thermodynamics and kinetics, self-assembly refers to the formation of ordered structures from individual building blocks or particles without direction from an external field. In recent years, the development of self-assembled structures from sub-micron-sized particle building blocks has attracted considerable attention due to their potential applications in nanotechnology \cite{li2010molecular,torring2011dna,wang2014electrostatic,li2015controlled,palchoudhury2017self,uchida2018modular,mccoy2018templated,wang2019controlling}.
An important challenge in elucidating and, eventually, engineering assembly pathways to optimize target structures is defining appropriate order parameters that quantify  local order in the assembling structures along the pathway\cite{wang2018nonlinear,spellings2018machine,adorf2019analysis,van2020classifying,coli2021artificial}. Defining suitable local order parameters is particularly challenging when the self-assembling structure is complex (e.g. a large unit cell, possessing chirality, etc.) or when competing polymorphs or pre-nucleation motifs emerge along the pathway \cite{teich2019identity, lee2019entropic, carpenter2021pre}. 

Previous studies have attempted to identify suitable order parameters for self-assembly, including using the radial distribution function, the bond-orientational order parameter \cite{mickel2013shortcomings,lechner2008accurate,coli2021artificial,mao2022deep}, and Voronoi tessellation \cite{stukowski2012structure,mickel2013shortcomings, larsen2016robust}. While these order parameters have been used extensively and successfully in capturing some aspects of the assembly process, they have limitations. For example, the radial distribution function (RDF) considers only the pairwise (two-point) correlation between particles and does not capture higher-order correlations. The bond-orientational order parameter is sensitive to local structure but is less effective in detecting global ordering. Methods using the Voronoi tessellation can elucidate local structure but are less effective in describing long-range order.  In particular, these approaches are insufficient for interrogating, e.g., the self-assembly of a system of truncated tetrahedra into a crystal with 432-particle unit cell \cite{lee2019entropic}.

Machine learning (ML) is becoming an increasingly popular approach for discovering order parameters useful in the study of self-assembly\cite{long2014nonlinear,reinhart2017machine,ferguson2017machine,spellings2018machine,defever2019generalized,adorf2019analysis,van2020classifying,reinhart2021unsupervised}. One ML approach uses existing order parameters that combine roto-translation-invariant symmetry functions as input descriptors and approximates optimal order parameters\cite{spellings2018machine,kim2020gcicenet}. However, an immediate shortcoming of this type of supervised ML method is its limited classification capabilities, which are constrained by its input descriptors. If none of the input descriptors can classify a certain crystal structure, the machine learning methods will fail to classify that crystal structure. Training deep neural networks (NNs) to classify crystal phases is another increasingly common  approach\cite{ziletti2018insightful,charest2020latent, coli2021artificial, mao2022deep, chakraborty2022deep, lizano2023convolutional}. Deep NNs can identify nonlinearity and are especially promising in constructing order parameters from particle features, such as instantaneous positions along a trajectory. Since thermal fluctuations often lead to noisy data, however, applying symmetry functions to encode particle positions is usually inevitable\cite{geiger2013neural}. These pre-engineered symmetry functions can work well for some systems, but finding symmetry functions that encode particle orientations is non-trivial. Finally, graph neural networks (GNNs) have also become popular for classifying crystal structures\cite{zou2023enhanced, dietrich2023machine, banik2023cegann}. By treating crystal structures as connected graphs, GNNs preserve permutation symmetry. However, they often require additional manipulation to deal with rotational and translational symmetry \cite{kim2020gcicenet}. Although equivariant GNNs have recently been proposed \cite{satorras2021n}, the question of how to incorporate equivariant properties for particle shapes remains to be addressed.  Common to all of these ML approaches is the tendency for increasing complexity in the approach as the building blocks and the structures they form become increasingly complex.

In this work, we show how we can use the most basic NN, a multilayer perceptron (MLP), as a local environment classifier in systems of particle shapes. Because we use the MLP classifier to quantify the local environment around each particle, the classifier is permutation-invariant, precluding the use of more sophisticated network structures such as GNNs\cite{banik2023cegann}. Thus, instead of employing conventional symmetry functions to transform per-particle quantities to input descriptors, we analyze step-by-step the symmetry of a particle's shape and propose a straightforward shape-symmetry encoded data augmentation method that allows our MLP classifier to operate on per-particle features directly. This data augmentation method ensures the roto-translation invariance of input features to the global environment and accounts for the particle shape's symmetry. Importantly, the simplicity of our data augmentation method necessitates minimal manipulation of the raw data, and also leads to a substantial improvement in the classification performance and flexibility when distinguishing different thermodynamic phases in our test systems. In this way, our approach can provide a simple, powerful, physics-agnostic alternative to conventional order parameters when studying self-assembly.

The remainder of this article is organized as follows. In Sec.~\ref{sec:in_features}, we introduce per-particle quantities as input features. Next, in Sec.~\ref{sec:data_aug}, we show how we perform data augmentation on input features based on particle shape symmetry. In Sec.~\ref{sec:mlp}, we introduce the MLP model we used as a local environment classifier. In Sec.~\ref{sec:results}, we demonstrate our method on seven different test cases. We first test our classifier's stability in Sec.~\ref{sec:simple_cubic} for a simple cubic structure self-assembled in simulation from hard cubes. Subsequently, we test our classifier on six additional self-assembly examples that result in three different categories of final products. In each case, we show how data augmentation improves classification quality. A conclusion is given in Sec.~\ref{sec:conclusion}.

\begin{figure*}[!t]
  \centering
  \includegraphics[width=0.9\textwidth]{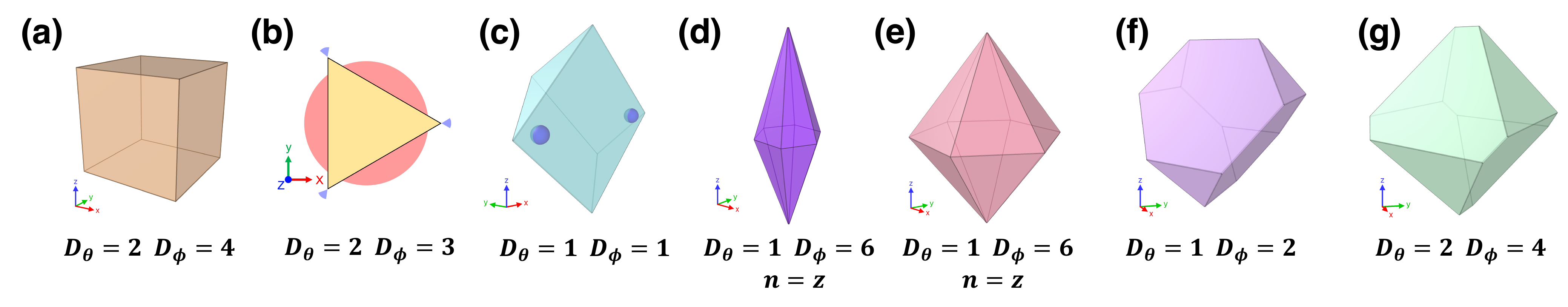}
  \caption
  {\textbf{Systems of particle shapes} The particle systems studied in this paper:  \textbf{(a)} cube, \textbf{(b)} patchy triangle, \textbf{(c)} patchy triangular prism\cite{zhou2022route},  \textbf{(d)} hexagonal bipyramid with aspect ratio $\alpha=3.0$, \textbf{(e)} hexagonal bipyramid with $\alpha=1.28$, \textbf{(f)} truncated tetrahedron, and \textbf{(g)} truncated octahedron. In (a) and (b), the transparent blue region decorating the particles indicates the attractive patch, while the transparent red region indicates a repulsive patch. Also shown are the shape symmetry-related factors $D_{\theta}$,  $D_{\phi}$, and mirror plane normal vector $\mathbf{n}$ utilized for each system. Each particle pictured here is oriented such that it is in the reference orientation given by the quaternion $\mathbf{q}_0=(1, 0, 0, 0)$.
}\label{fig:particle_sys} 
\end{figure*}

\section{Method}
\label{sec:method}
We construct a classifier for the local environment around a particle using an MLP that incorporates continuous, discrete and shape symmetry information. By building an MLP classifier that learns to classify each particle, we ensure that the classification is permutation invariant.  We first extract features from each particle's local environment and then use these features to train MLP classifiers to accurately classify particles based on their local environment.

\begin{figure}[!t]
  \centering
  \includegraphics[width=8cm]{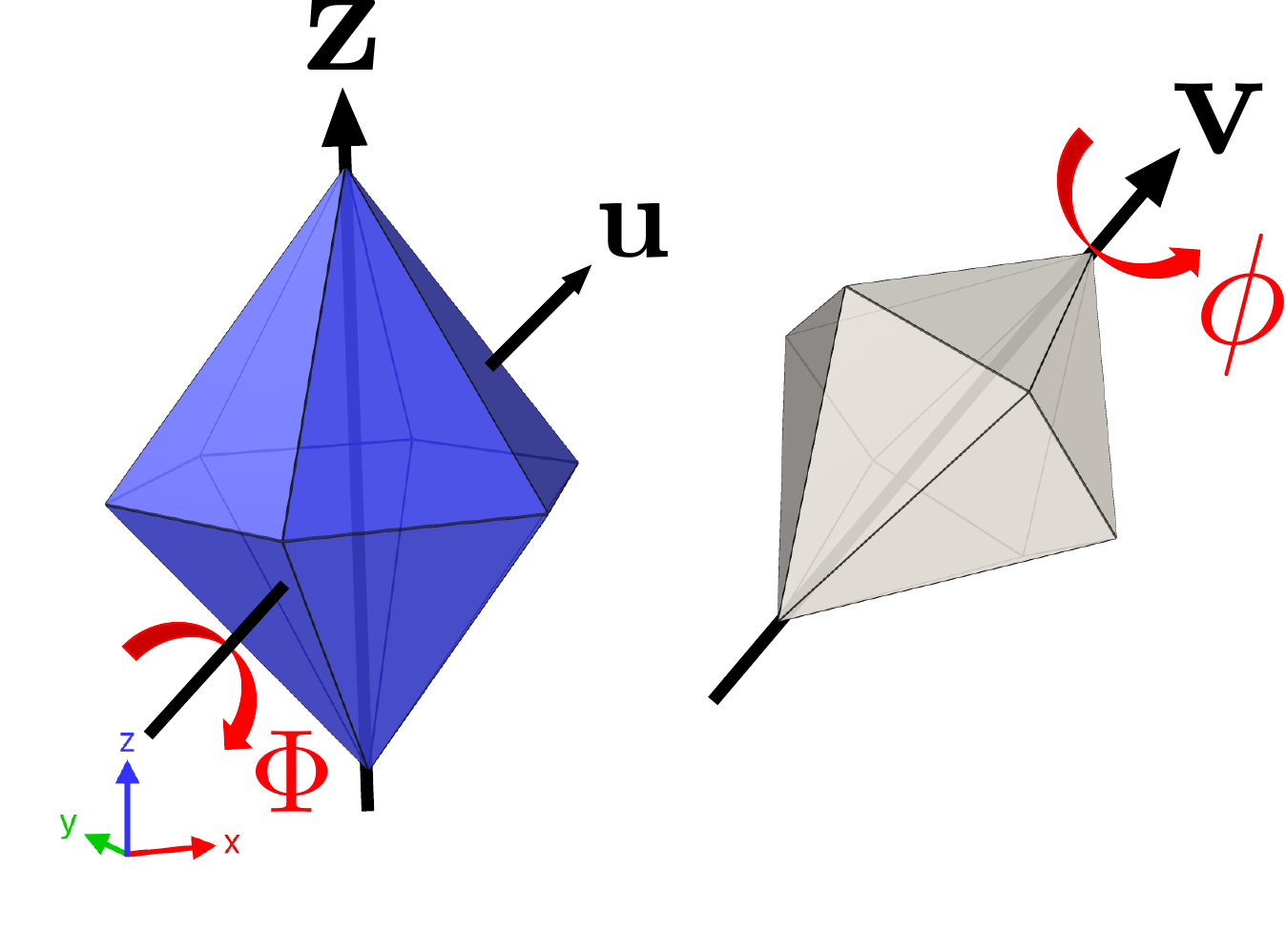}
  \caption
  {\textbf{Rotation in the local environment} The blue bipyramid represents the target particle, and the white bipyramid represents a neighboring particle.
  }\label{fig:rotation}
\end{figure}

\subsection{Per-particle quantities as input features}
\label{sec:in_features}
To account for translational symmetry and describe the local environment of a target particle in a system of like particles, we employ a set of interparticle quantities defined with respect to the surrounding neighborhood. Specifically, we calculate the \textbf{relative positions} $\mathbf{r}_{ij}$ and \textbf{relative orientations} defined by quaternion $\mathbf{q}_{ij}$ of a predetermined number of neighboring particles $j$ in relation to the target particle $i$. These quantities serve as a unique fingerprint of the local environment surrounding the target particle \cite{keys2011characterizing, coli2021artificial} and are defined below. 

We define the relative position $\mathbf{r}_{ij}$ using spherical coordinates $(r_{ij}, \theta_{ij}, \phi_{ij})$, where the relative distances $r_{ij}$ between a fixed number $N_b$ of neighboring particles denoted by the index $j$ (that is, the size of the neighborhood) and the target particle $i$ contain information about the local density distribution. The angular part $(\theta_{ij}, \phi_{ij})$ define the local bond angles.
Note that for a two-dimensional system, $\theta_{ij}$ is always 0. However, despite the invariance of the relative distances $r_{ij}$ to arbitrary translation operations, these distances can still be affected by thermal fluctuations and the length scale of the system, which can impact the transferability of our classifier. Thus, we normalized the relative distances:
\begin{align}
    r_{ij}^{\mathrm{(au)}} = \frac{r_{ij}}{\max_{j \in N_b} r_{ij}}
    \label{eq:r_ij}
\end{align}
\noindent where the superscript $^\mathrm{(au)}$ denotes the augmented features. \\

The relative quaternion $\mathbf{q}_{ij}$ is commonly used to define the relative orientation of a neighboring particle $j$ with respect to the target particle $i$. The quaternion plays an important role in both simulating and analyzing crystallization of a system of anisotropic particles \cite{haji2015strong, ramasubramani2018rowan} and in defining the space group of crystals of certain molecular systems. To describe a particle's orientation, a reference orientation first needs to be established  using the important symmetry axis of the particle along the orthogonal basis of Cartesian coordinates. We denote the reference orientation by $\mathbf{q}_0 = (1, 0, 0, 0)$, which is equivalent to the identity rotation with respect to the reference orientation. In Fig.~\ref{fig:particle_sys}, we illustrate the particles studied in this paper by placing them in the predefined reference orientation. Given this reference orientation, we can define subsequent orientations of each particle by performing a 3-dimensional spatial rotation, which can be expressed as a rotation quaternion $\mathbf{q} = (C, Su^x, Su^y, Su^z)$ where $(C, S) = (\cos{(\theta/2)}, \sin{(\theta/2)})$. This $\mathbf{q}$ thus represents a rotation angle $\theta$ from the reference orientation about the axis $\mathbf{u}$. The relative orientation between target particle $i$ and its neighborhood $j$ can then be expressed as the rotation quaternion via the conventional rotation from $i$ to $j$ as:

\begin{align}
    \mathbf{q}_{ij} = \mathbf{q}_{i}^{-1} \mathbf{q}_{j}
    \label{eq:q_ij}
\end{align}

By utilizing relative position and orientation, we can accurately capture each particle's local environment while maintaining translational invariance. We have yet to discuss the property of invariance to an arbitrary rotation of the system, which can influence the angular parts $\theta_{ij}, \phi_{ij}$ that are used to define relative position and $\mathbf{q}_{ij}$. Conventionally, the rotational invariance is achieved by randomly rotating the training set in each training epoch\cite{dietrich2023machine}; this step is referred to as data augmentation. In the next section, we present a more robust approach for training a classifier invariant to an arbitrary system rotation of particle environments.

\subsection{Shape-symmetry encoded data augmentation}
\label{sec:data_aug}
In the previous section, we introduced relative positions and quaternions as input features. These input features are designed so they are invariant to the translation of the  system. However, arbitrary rotation of the system can also limit the transferability of the classifier, which is traditionally addressed using data augmentation techniques\cite{defever2019generalized}. For example, in image processing, one duplicates the image but with random rotations to create a training dataset to prevent possible overfitting and enhance the transferability of the classifier. Recent developments in equivariant NNs (ENNs) also allow for input without data augmentation\cite{satorras2021n,finzi2021practical,chen2022robust}. For systems of point particles, the final crystal structure can be used to define the reference orientation. Here we exploit particle shape and use the user-defined reference orientation represented by quaternions to define the local environment. To ensure rotational invariance, we independently rotate each target particle and its local environment onto the predefined reference orientation, such that each target particle is in the predefined reference orientation as shown in Fig.~\ref{fig:particle_sys} prior to calculating the per-particle quantities.

In addition to translational and rotational symmetries associated with the system of particles, the individual particles can possess symmetries. With a sophisticated design, the ENNs can process these additional symmetries during training. However, we propose a simple yet effective way to achieve the same result here. We perform an additional data augmentation that encodes the particle's shape and interaction anisotropy (patchiness). The new angular part of the relative position and relative quaternion after this data augmentation is denoted as $(\theta_{ij}^{\mathrm{(au)}}, \phi_{ij}^{\mathrm{(au)}})$ and $\mathbf{q}_{ij}^{\mathrm{(au)}}$. Through simple geometric reasoning, the augmented angular part can be calculated easily as follows:
\begin{align}
    (\theta_{ij}^{\mathrm{(au)}}, \phi_{ij}^{\mathrm{(au)}})= \left(
    \frac{\theta_{ij}}{\pi}\bmod\frac{1}{D_{\theta}}, \frac{\phi_{ij}}{\pi}\bmod\frac{2}{D_{\phi}} \right)
    \label{eq:augmented_angle}
\end{align}

\noindent where $\bmod$ is the modulo operator, and $D_{\phi}$ and $D_{\theta}$ are discrete integers that assume the value, $N$ of the $N$-fold rotational symmetry along $\phi$ and $\theta$-direction with respect to the particles' shape when it is in the reference orientation. In Fig.~\ref{fig:particle_sys}, we show $D_{\phi}$ and $D_{\theta}$ for each of the corresponding hard shapes and patchy particles in their reference orientations.

The orientation quaternion $\mathbf{q}_{ij}$ ican be written as a rotation of angle $\Phi$ along an axis $\mathbf{u}$, as shown in Fig.~\ref{fig:rotation} It can be proved that the same quaternion can also be written as the rotation of angle $\phi$ along axis $\mathbf{v}$, 

\begin{align}
\mathbf{q}_{ij}=e^{i\frac{\Phi}{2}\mathbf{u}}=e^{i\frac{\phi}{2}\mathbf{v}}\mathbf{Q}
\label{eq:non_augmented_q}
\end{align}

\noindent where $\mathbf{v}=\mathbf{Q}\mathbf{z}\mathbf{Q}^{-1}$ defines the symmetry axis of the particle. The symmetry axis is defined as the z-axis when each particle is in its reference orientation and rotates in the same way the particle rotates. The augmented quaternion is then calculated as
\begin{align}
    \mathbf{q}_{ij}^{\mathrm{(au)}}=e^{i\frac{\phi^{\mathrm{(au})}}{2}\mathbf{v}^{\mathrm{(au)}}}
    \label{eq:augmented_q}
\end{align}

\noindent where $\phi^{\mathrm{(au)}}=\phi\bmod\frac{2\pi}{D_{\phi}}$ and

\begin{align}
    \mathbf{v}^{\mathrm{(au)}}=\mathbf{v}_{\perp\mathbf{z}}+\cos\theta^{\mathrm{(au)}}\mathbf{z}
\label{eq:augmented_v}
\end{align}

\noindent where $\mathbf{v}_{\perp\mathbf{z}}$ is defined as the component perpendicular to $\mathbf{z}$ and $\theta^{\mathrm{(au)}}=\arccos(\mathbf{v}\cdot\mathbf{z})\mod\frac{\pi}{D_{\theta}}$.

Under the same rule, it is also straightforward to consider mirror symmetry. For the angular part, in general, we can calculate augmented relative positions $\mathbf{r}_{ij}^{\mathrm{(au)}}$ before separating relative positions into distance $r_{ij}$ and angular parts $(\theta_{ij}, \phi_{ij})$:
\begin{align}
    \mathbf{r}_{ij}^{\mathrm{(au)}}=(\mathbf{r}_{ij})_{\perp\mathbf{n}} + \|\mathbf{r}_{ij}\cdot\mathbf{n}\|\mathbf{n}
    \label{eq:augmented_position}
\end{align}

\noindent where $(\mathbf{r}_{ij})_{\perp\mathbf{n}}=\mathbf{n}\times(\mathbf{r}_{ij}\times\mathbf{n})$ is defined as the component perpendicular to $\mathbf{n}$. In practice, since the mirror plane of a hexagonal bipyramid is along the $\mathbf{z}$-axis, we need only to take the absolute value of the $\mathbf{z}$-coordinate before separating relative positions into distance and angular parts. For the quaternion part, we need to separate the symmetry axis into normal and parallel parts with respect to the plane normal vector $\mathbf{n}$:
\begin{align}
    \mathbf{v}^{\mathrm{(au)}}=\mathbf{v}_{\perp\mathbf{n}}+\|\mathbf{v}\cdot\mathbf{n}\|\mathbf{n}
\label{eq:augmented_v}
\end{align}
where $\mathbf{v}_{\perp\mathbf{n}}=\mathbf{n}\times(\mathbf{v}\times\mathbf{n})$ is defined as the component perpendicular to $\mathbf{n}$.

\subsection{Multilayer Perceptron (MLP) as local environment classifier}
\label{sec:mlp}

\begin{figure}[!t]
  \centering
  \includegraphics[width=8cm]{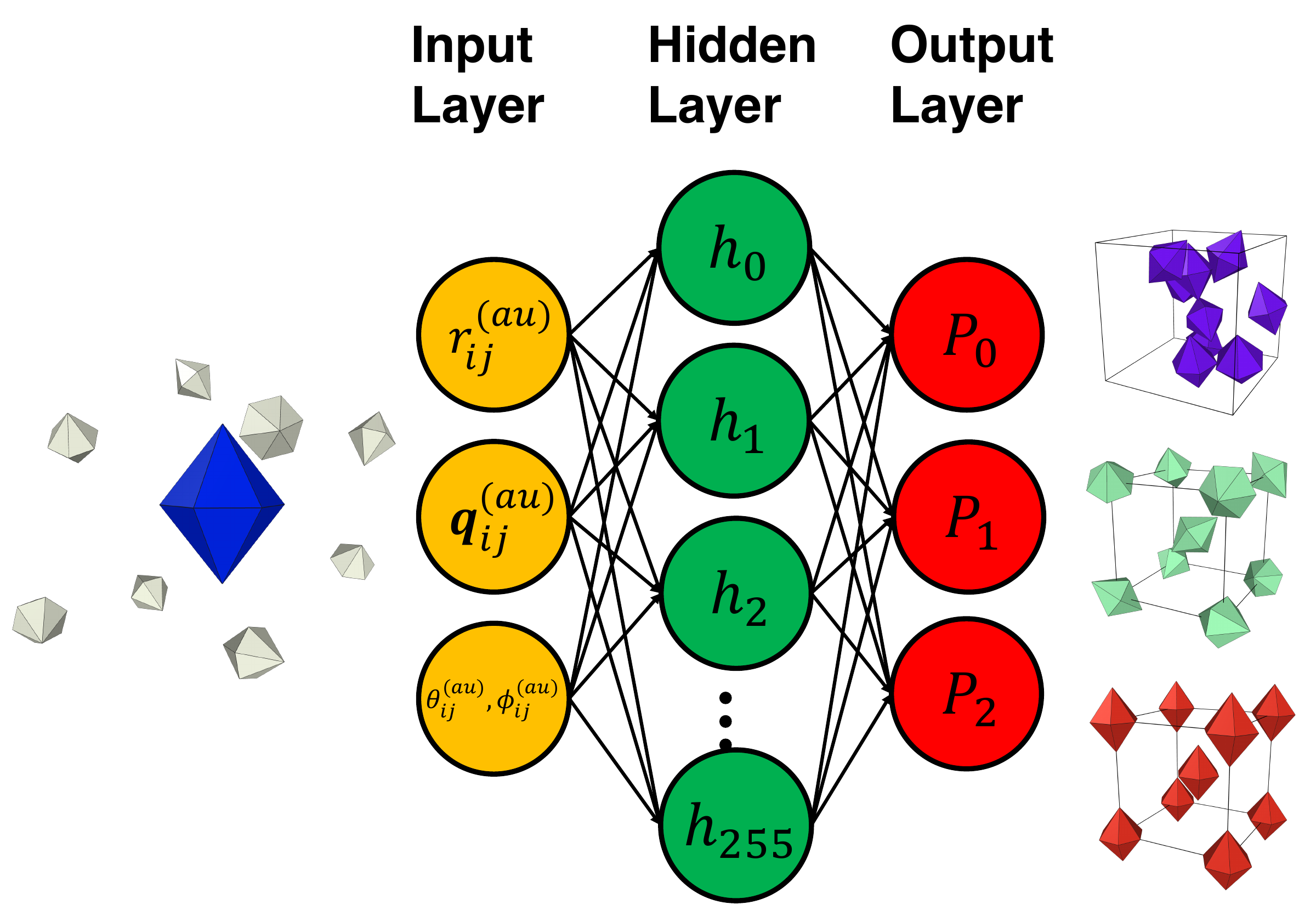}
  \caption
  {\textbf{The Architecture of Multilayer Perceptron (MLP)} The diagram of the MLP architecture used to classify local particle environment. The input layer has $N_b + 4N_b + 2N_b$ units as described in Eq.~\ref{eq:x_i}, and the hidden layer contains 256 units. The number of output layers depends on the system. For example, we will present results for a system of particles undergoing a two-steps crystallization transition, where the output layer is composed of three units representing the probability that a given particle belongs to one of three predefined phases.
  }\label{fig:mlp}
\end{figure}

The local symmetry is broken during a self-assembly process as the local environment around a particle changes. To monitor the change in the local environment, we utilize a fully connected NN, commonly known as a multilayer perceptron (MLP), to classify this local environment, as illustrated in Fig.~\ref{fig:mlp}.  There are several advantages of using a simple MLP instead of a more advanced GNNs or ENNs. First, an MLP handles large datasets with simplicity and effectiveness. An MLP allows us to train the machine even on a personal laptop, suitable for a quick, on-the-fly test. Additionally, an MLP can be easily extended to incorporate additional features or classify particles in other colloidal systems. Second, a simple MLP can be greatly accelerated by harnessing the power of modern graphical processing units (GPUs), which can facilitate future research, such as using MLPs as order parameters to study the assembly pathways of disparate systems forming the same structure, or in enhanced sampling methods such as umbrella sampling or metadynamics whose algorithms demand efficient calculation of order parameters and their derivatives. Third, despite its simple network structure, a MLP still provides sufficient nonlinearity to build a powerful classifier from fundamental features, e.g., particle coordinates and orientations. Its  classification ability is decent enough to map local environmental fingerprints of particle shapes to thermodynamic phases.

The input to the MLP classifier is a set of feature vectors $(r_{ij}^{\mathrm{(au)}}, \mathbf{q}_{ij}^{\mathrm{(au)}}, \theta_{ij}^{\mathrm{(au)}}, \phi_{ij}^{\mathrm{(au)}})$, as described in Sec.~\ref{sec:in_features} and ~\ref{sec:data_aug}. These feature vectors are arranged and concatenated to a 1-dimensional vector $\mathbf{x}_i$ with the relative position and relative quaternion sequentially placed in ascending order as follows,
\begin{align}
  \mathbf{x}_i = \mathrm{sort}_{r_{ij}^{\mathrm{(au)}}}(..., r_{ij}^{\mathrm{(au)}}, 
  \mathbf{q}_{ij}^{\mathrm{(au)}}, 
  \theta_{ij}^{\mathrm{(au)}},\; \phi_{ij}^{\mathrm{(au)}},\; ...)
  \label{eq:x_i}
\end{align}

\noindent where $j$ runs over all neighbors and therefore $\mathbf{x}_i \in \mathbb{R}^{N_b + 4N_b + 2N_b}$. As shown in Fig. \ref{fig:mlp}, the MLP architecture includes one hidden layer consisting of 256 neurons. The input layer of the network takes the feature vectors $\mathbf{x}_i$ and propagates them forward to the hidden layer, which performs a linear operation followed by a non-linear activation function $\sigma$:

\begin{align}
  \mathbf{h} = \sigma(\mathbf{W}\mathbf{x}_i+\mathbf{b})
\label{eq:mlp_layers}
\end{align}

\noindent The activation function comprises a linear operation defined by a layer-wise matrix multiplication with trainable weight matrix $\mathbf{W}$ and bias vector $\mathbf{b}$.

The hidden layer $l=0$ takes input feature vectors $\mathbf{x}_i$ directly, and therefore has dimensions $\mathbf{W} \in \mathbb{R}^{\mathcal{M} \times \mathcal{N}}$ and $\mathbf{b} \in \mathbb{R}^{\mathcal{M}}$, where $\mathcal{M}=256$ and $\mathcal{N}=N_b + 4N_b + 2N_b$. After this linear operation, we used a rectified linear unit (ReLU) as the activation function $\sigma$.   The output from the hidden layer is then converted to output nodes $\mathbf{y}\in\mathbb{R}^{\mathcal{C}}$ in the output layer,

\begin{align}
\hat{\mathbf{y}}=\mathrm{softmax}(\mathbf{W}^{(o)}\mathbf{h}+\mathbf{b}^{(o)})
\label{eq:mlp_output}
\end{align}

\noindent where $\mathbf{W}^{(o)}\in\mathbb{R}^{\mathcal{M}\times\mathcal{C}}$ and $\mathbf{b}^{(o)}\in\mathbb{R}^{\mathcal{C}}$ are the output weight matrix and output bias, respectively, and $\mathcal{C}$ is the number of pre-defined classes. The $\mathrm{softmax}$ function in eq. \ref{eq:mlp_output} is a mathematical function that converts a vector of real numbers into a probability distribution. The MLP classifiers are then trained using the optimizer Adam\cite{KingBa15}, with a learning rate of $7.5\times 10^{-4}$, and with an error metric given by the cross-entropy loss $L$, defined as:
\begin{align}
    L=-\sum^{\mathcal{C}}_{n=1}\log\hat{\mathbf{y}}_n
    \label{eq:cel_loss}
\end{align}

\noindent The cross-entropy loss $L$, or log loss, measures the performance of a probabilistic classification model whose output is a probability distribution. For simplicity, all of our MLP classifiers are trained with 30 epochs, where the number of epochs is defined as the number of times the optimization algorithm goes through all training samples. The classifier and training algorithm are both implemented using PyTorch\cite{NEURIPS2019_9015}.

\subsection{Data preparation}
\label{sec: training_set}
We generated independent training and testing trajectories for all seven test systems. Our training trajectories comprise fully equilibrated phases that were initiated from synthetic structures or self-assembled and annealed at various thermodynamic conditions. Equally spaced snapshots of the training trajectories were used to generate the training sets, from which we randomly drew the validation sets. The partitioning ratio of training and validation sets was fixed at 4:1, and each training set comprised a minimum of 20,000 local environments. The testing trajectories comprised self-assembly runs in which at least one phase transition was observed in each run.

We utilized the Hard Particle Monte Carlo (HPMC) and Molecular Dynamics (MD) modules of HOOMD-Blue\cite{anderson2020hoomd} to simulate two-dimensional and three-dimensional convex hard particles. Interaction patchiness was implemented using the Just-In-Time (JIT) compilation module under HPMC. We simulated the equilibration, annealing and self-assembly processes within both the canonical (NVT) and isobaric-isothermal (NPT) ensemble. To simulate the hard particles using MD, we employed the anisotropic Weeks-Chandler-Andersen (AWCA) potential in HOOMD-Blue\cite{ramasubramani2020mean}.

After we obtained the trajectories, we used the freud analysis package\cite{ramasubramani2020freud} to construct the neighbor list used to calculate per-particle input features. Other analysis functions of freud were also used, such as the radial distribution function (RDF) and Steinhardt order parameters\cite{steinhardt1983bond}. Snapshot images are rendered using Ovito\cite{stukowski2009visualization}.

Each test case was simulated as follows:

\begin{description}
    \item[$\bullet$ Cubes] For training trajectories, three independent equilibrated HPMC training trajectories were prepared, with one dense fluid trajectory and two crystal trajectories equilibrated at packing fractions of 0.244 and 0.751, respectively. The testing trajectory was generated by compressing the system from packing fraction 0.244 to 0.864 using HPMC simulation and then equilibrating.
    
    \item[$\bullet$ Patchy triangles] For training trajectories, three independent equilibrated training HPMC trajectories were prepared, with one dense fluid trajectory and two kagome lattice trajectories. The kagome lattice is a two-dimensional crystal composed of corner-sharing triangles that has been discovered to be assembled by triblock Janus particles\cite{chen2011directed, rivera2023inverse}. The fluid phase was equilibrated above the nucleation temperature ($k_BT \gtrapprox 0.105$), while the kagome lattice was initialized from a perfect kagome lattice with randomly placed guest particles and equilibrated at $k_BT =0.105$.  The testing trajectory was prepared by quenching from $k_BT=0.3$ to 0.1 and equilibrating using HPMC.

    \item[$\bullet$ Patchy prisms] For training trajectories, three independent equilibrated HPMC training trajectories were prepared, with one dense fluid trajectory and two crystal trajectories. The testing trajectory was generated by equilibrating an initially disordered system. Here, we use the system from Ref.~\cite{zhou2022route} where the distance between the attractive patches and the face center of the prism is 0.8.

    \item[$\bullet$ Hexagonal bipyramids with aspect ratio $\alpha = 1.28$] For training trajectories, three independent equilibrated training MD trajectories were prepared, with one dense fluid trajectory, one plastic crystal trajectory, and one body-centered tetragonal (BCT) crystal trajectory\cite{lim2023engineering}. The systems were equilibrated at packing fractions 0.464 and 0.569 for the dense fluid and plastic crystal, respectively.

    \item[$\bullet$ Hexagonal bipyramids with aspect ratio $\alpha = 3.0$] For training trajectories, three independent equilibrated training MD trajectories were prepared, with one dense fluid trajectory, one liquid crystal trajectory, and one triclinic crystal trajectory \cite{lim2023engineering} equilibrated at packing fractions of 0.4, 0.51, and 0.661, respectively. The testing trajectory was prepared by quenching the system from a reduced pressure $P^*$ of 0.5 to 10 and equilibrating it using MD.

    \item[$\bullet$ Truncated tetrahedrons and octahedrons] For truncated tetrahedrons,  three independent HPMC training trajectories were prepared, with one dense fluid trajectory and two crystal trajectories. The dense fluid trajectories were equilibrated at packing fractions 0.347 for truncated tetrahedrons and 0.524 for truncated octahedrons. For truncated tetrahedrons, the two training diamond crystal trajectories were equilibrated at packing fractions of 0.561. For the truncated octahedrons, the two training high-pressure lithium crystal trajectories were prepared by slowly annealing a self-assembled crystal and equilibrating at a packing fraction of 0.606. The HPMC testing trajectories were prepared by quenching the systems from a reduced pressure $P^*$ from 0.5 to 10 for truncated tetrahedrons (using NPT) and a packing fraction from 0.14 to 0.62 for truncated octahedrons (using NVT), followed by equilibration. 
\end{description}

\section{Results and Discussion}
\label{sec:results}
We test the performance of our MLP on seven different systems, beginning with the simplest case of hard cubes that self-assemble into a simple cubic lattice.

\subsection{Test case 1: Simple cubic crystals assembled by hard cubes}
\label{sec:simple_cubic}

\begin{figure*}[!t]
  \centering
  \includegraphics[width=0.9\textwidth]{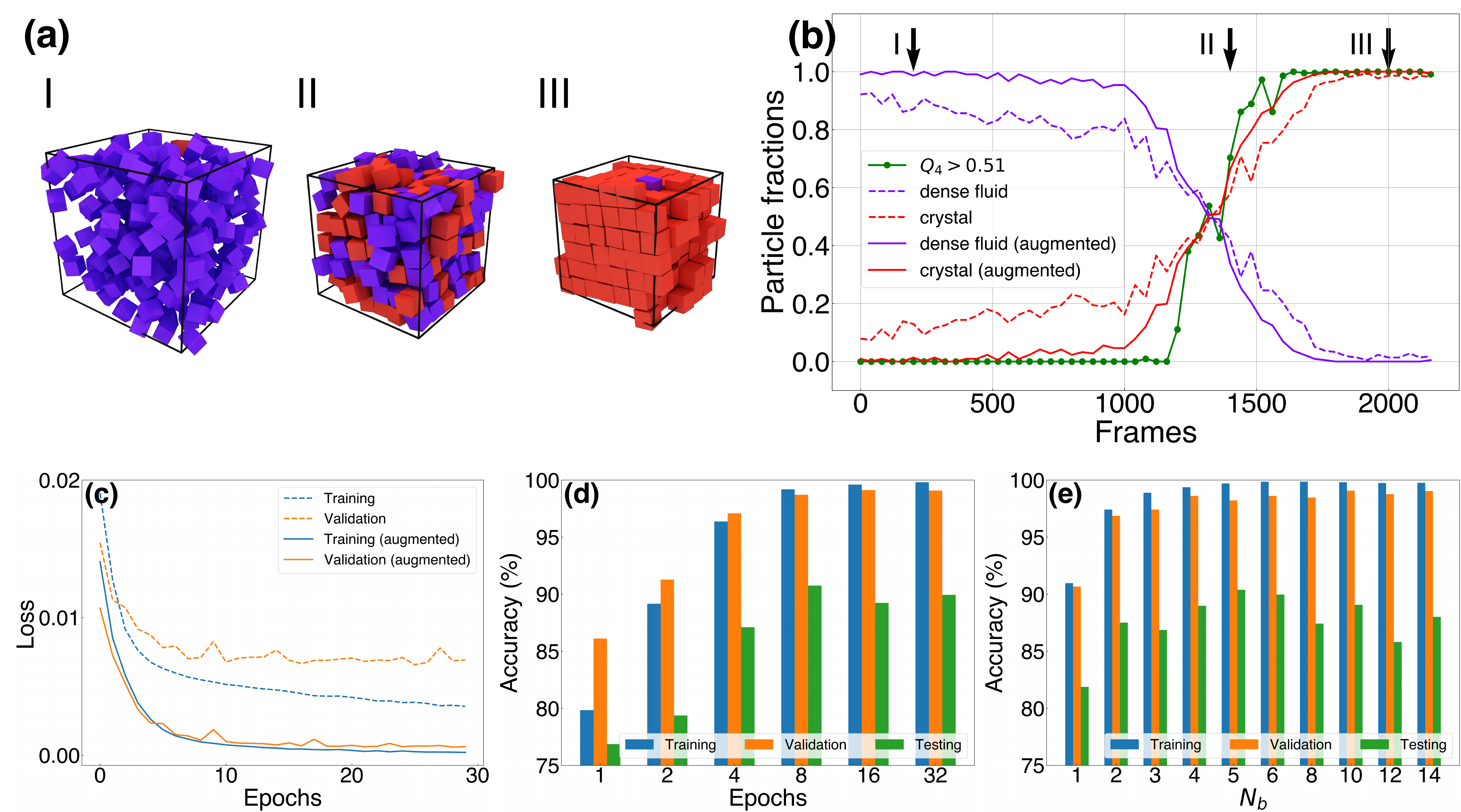}
  \caption
  {\textbf{Simple cubic lattice consisted of cubes}. Summary of the MLP classifier's classification results, training details, and accuracy tests, in which we use the simple cubic system as a test case. \textbf{(a)} The MLP classifier's classification results on three different snapshots. \textbf{(b)} The MLP classifier's classification results on the self-assembly trajectory and classification, compared to the Steinhardt order parameter $Q_4$ (dotted line) calculated as ground truth. For visualization purposes, solid lines represent the MLP classifier trained on the data accounting for symmetry, while dashed lines are for the MLP classifier trained on the data without accounting for symmetry. Annotations I, II, and III indicate the three corresponding snapshots in (a) for the classifier trained on augmented data. \textbf{(c)} Learning curve of the MLP classifier used in (a) with (solid line) and without (dashed line) data augmentation \textbf{(d)} MLP classification accuracy plotted versus increasing number of training epochs \textbf{(e)} MLP classification accuracy plotted versus increasing number of neighbors $N_b$ in the local particle neighborhood. In (a)-(d), we used fixed $N_b = 6$. In (a)-(c) and (e), we used 30 training epochs.
}\label{fig:cube}
\end{figure*}

As a demonstration of our method, we first show a simple classification test on a simple cubic structure self-assembled from the fluid phase of hard cubes upon an  increase in packing fraction. As shown in Fig.~\ref{fig:particle_sys}, the cube exhibits 2-fold and 4-fold rotational symmetries along the $\theta$- and $\phi$-directions. Using Eq.~\ref{eq:augmented_angle} and Eq.~\ref{eq:augmented_q}, we can explicitly consider these symmetries in the data augmentation step and generate an appropriate training dataset that accounts for symmetry. From the three snapshots, each representing a different stage in the self-assembly process, in Fig.~\ref{fig:cube} (a), we see that as the packing fraction increases, the number of local environments classified as locally cubic also increases.

To further quantify the extent to which data augmentation using symmetry improves the MLP's classification abilities, we compare the classification results by the classifier trained with and without data augmentation. In Fig.~\ref{fig:cube} (b), the particle fraction is defined as the number of particles being classified as a certain local environment divided by the total number of particles in the system. There, we also plot the conventional Steinhardt order parameter $Q_l$\cite{steinhardt1983bond} with $l=4$ for comparison. The order parameter $Q_l$ is defined by averaging $Q_{l, i}$ over each particle $i$ in a system defined as:
\begin{align}
Q_{l, i} &= \sqrt{\frac{4 \pi}{2l+1}\sum_{m=-l}^{l}Q_{lm, i}Q_{lm, i}^{*}} \\
Q_{lm, i} &= \frac{1}{N_b}\sum_{j=1}^{N_b} Y_{lm,\; ij} (\mathbf{r}_{ij})
\label{eq:steinhardt}
\end{align}

\noindent where $Y_{lm,\; ij}$ is the spherical harmonic calculated by the relative position of the target particle $i$ and its neighbors $j$. For clarification, $Q_l$ differs from the quaternion $\mathbf{Q}$ in equation \ref{eq:non_augmented_q}. 

Note that when using $Q_4$, we need to select a threshold value (manually selected as 0.51 in this case) based on human intuition or visualization, or we can apply another ML method to determine this threshold, such as support vector machine (SVM), which maximizes the margin. While in the MLP classifier, the threshold is determined solely by the MLP. It can be clearly seen that without data augmentation based on symmetry, misclassification occurs primarily in the fluid phase. In Fig.~\ref{fig:cube} (c), we see from the convergence behavior of the loss values that the augmented dataset generally converges faster and better.

Fig.~\ref{fig:cube} (d) and (e) shows the results of accuracy tests performed on two other aspects. First, in Fig.~\ref{fig:cube}(d), we observe that accuracy increases with the number of training epochs. We observe that the accuracy converges at around eight epochs and persists without severe overfitting until 32 epochs. Second, in Fig.~\ref{fig:cube} (e), we show the change in accuracy by increasing $N_b$. The accuracy converges after $N_b=5$, roughly the number of first nearest neighbors in a cubic lattice.  This provides an excellent initial guess for $N_b$ in preparing the training set of our classifier. These tests show that our classifier is robust with appropriate augmentation, training epochs, and $N_b$. In the next section, we will look at more complicated crystals formed by hard particle shapes with different symmetries.

\subsection{Test cases 2 and 3: Self-assembly of 2D and 3D patchy particles}
\label{sec:kagome}

\begin{figure*}[!t]
  \centering
  \includegraphics[width=0.9\textwidth]{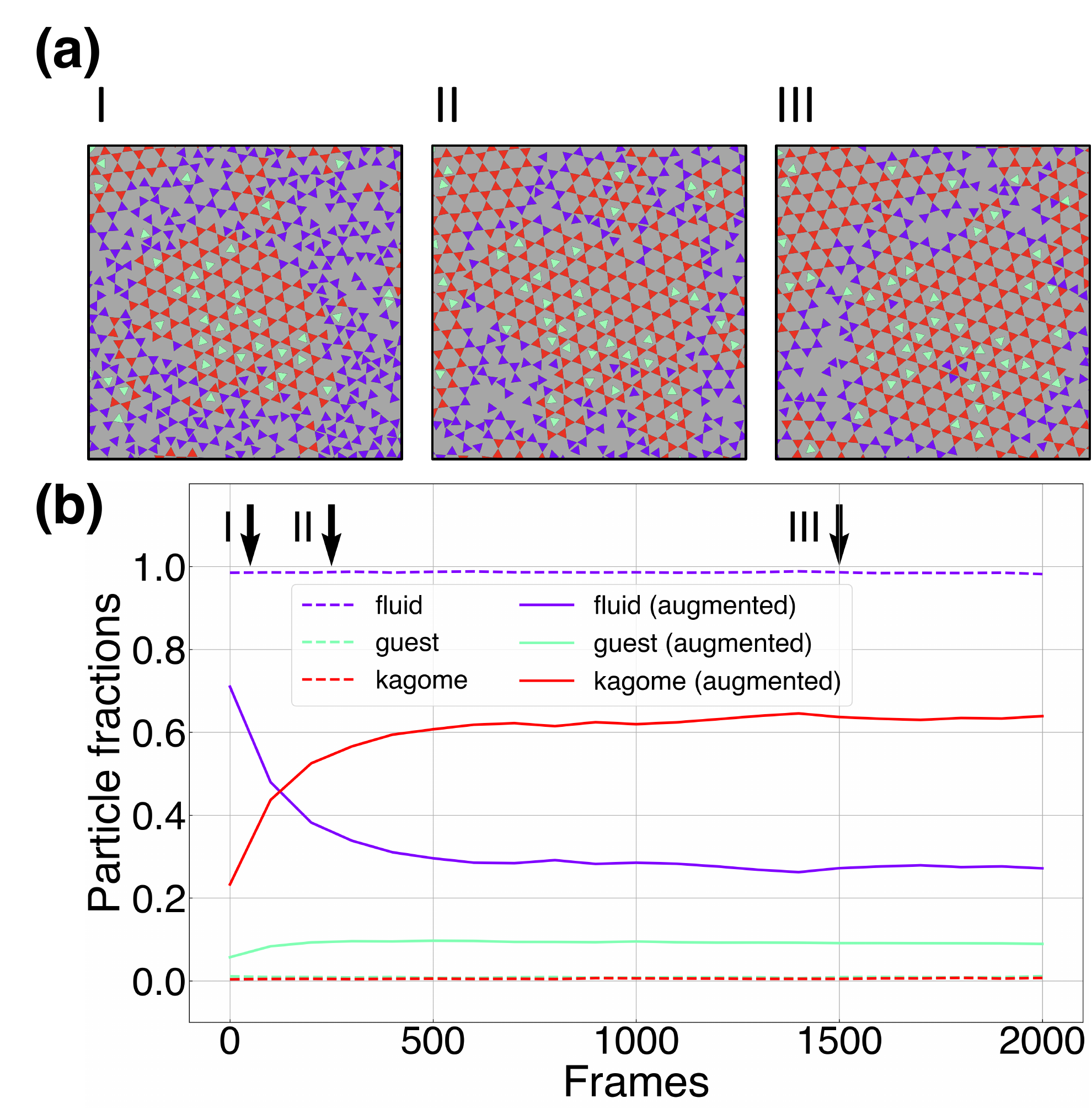}
  \caption
  {\textbf{Kagome lattice of patchy triangles} Summary of the MLP classifier's classification results on the test assembly trajectory. \textbf{(a)} The MLP classifier's classification results on the three snapshots. \textbf{(b)} The MLP classifier's classification results on the entire trajectory. For visualization purposes, solid and dashed lines represent the MLP classifier trained on the data with and without augmentation, respectively. The annotations I, II, and III correspond to the three snapshots in (a) for the classifier trained on augmented data. 
}\label{fig:kagome}
\end{figure*}

For this second test case, we employ our classifier to classify 2D and 3D systems of patchy particles. In the 2D case depicted in Fig.~\ref{fig:kagome} (a), each particle is a rigid equilateral triangle (Fig.~\ref{fig:particle_sys} (b)). The patchiness is realized by decorating each particle with a Kern-Frenkel attractive patch\cite{kern2003fluid} at each of the three vertices. Additionally, we apply three repulsive patches centered on each of the particle's edges to negatively design against undesirable phases. The guest particles inside the kagome lattice make finding a reliable order parameter that distinguishes the guest particles from non-guest kagome particles necessary. Our simulation results show that during self-assembly three distinct local environments emerge corresponding to a fluid-like, guest particle, and kagome lattice environments . 

We demonstrate the classification result of our MLP classifier on the test case assembly pathway of the kagome lattice in Figure \ref{fig:kagome} (b), where the light green color indicates the local environment is classified as a guest particle. It can be seen that the system nucleates and forms a kagome lattice cluster within the fluid phase. Inside the kagome lattice cluster, several guest triangles are enclosed by six surrounding triangles. As the assembly simulation proceeds, we observe the coalescence of  multiple small clusters into a single crystallite. Once the majority of particles in the system are in kagome lattice phase, the number of guest particles remains unchanged, which is also captured by the MLP classifier. It should be noted that some guest particles are misclassified as belonging to the fluid phase when they are too close to the surrounding particles within the kagome lattice. On the other hand, without data augmentation, the classifier only discovered fluid phase. This inability to distinguish local environments without data augmentation is consistent with our observation in hard cubes that when the crystal phase forms, the MLP classifier without data augmentation tends to underestimate the local environments of the ordered phases.

\begin{figure*}[!t]
  \centering
  \includegraphics[width=0.9\textwidth]{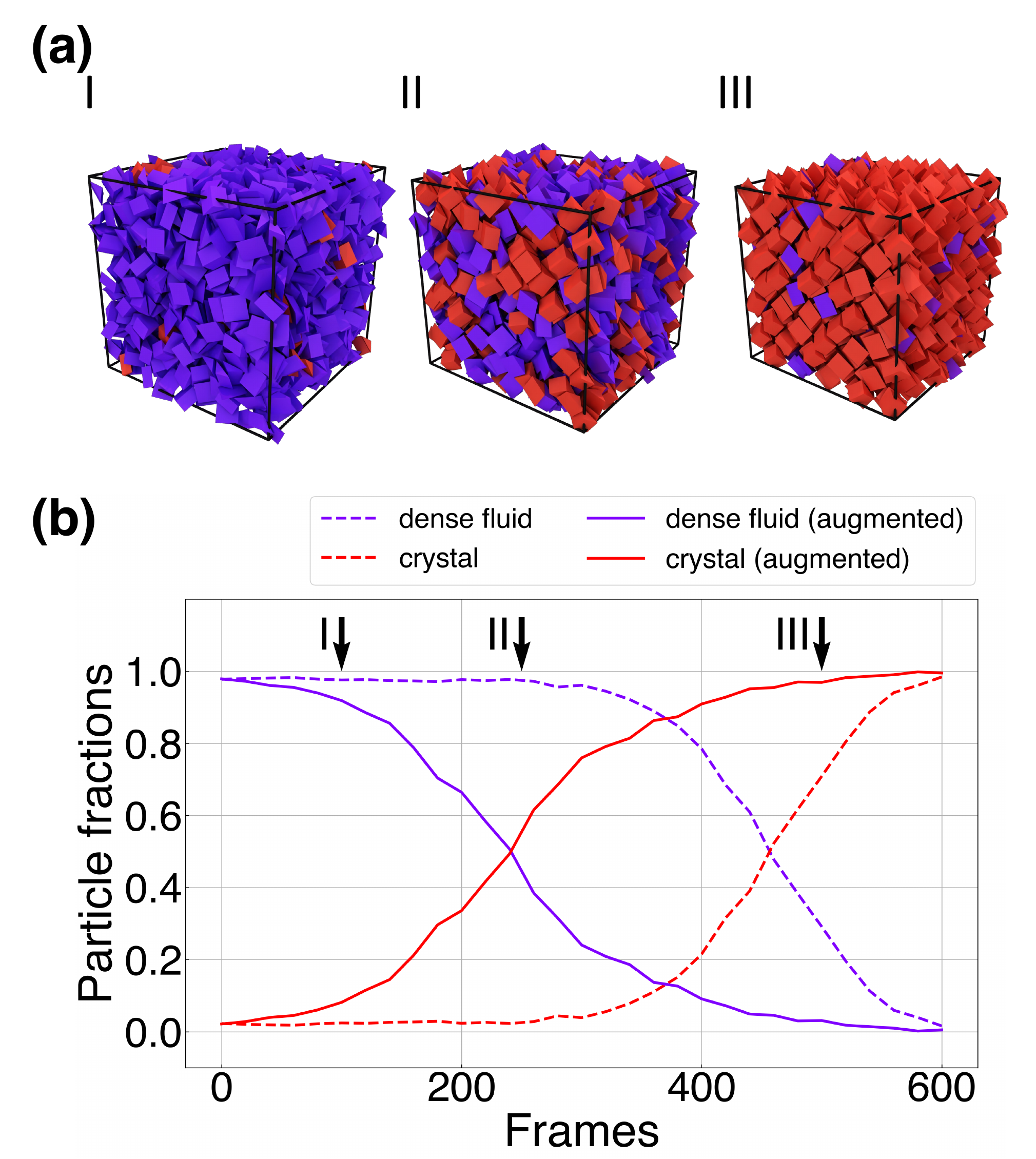}
  \caption
  {\textbf{Dimer-diamond phase consisted of anti-aligned patchy triangular prism\cite{zhou2022route}} The MLP classifier's classification results on the testing self-assembly trajectory, in which we identify the crystallization dimer-diamond structure. Each Wyckoff site comprises two anti-aligned patchy prisms forming a gyrobifastigium.   
  \textbf{(a)} The MLP classifier's classification results on the three snapshots. \textbf{(b)} The MLP classifier's classification results on the whole trajectory. For visualization purposes, solid and dashed lines represent the MLP classifier trained on the data with and without augmentation, respectively. The annotates I, II, and III indicate the corresponding snapshots in (a) for the classifier trained on augmented data.
}\label{fig:dimer_diamond}
\end{figure*}

The second test case is the 3-dimensional dimer diamond self-assembled from the fluid phase of patchy triangular prisms\cite{zhou2022route} (See Fig.~\ref{fig:particle_sys} (c)). In this test example, the triangular prisms pair up and arrange the pairs into the dimer-diamond structure. Since we treat each triangular prism as a particle, it is insufficient for the MLP classifier to classify only the diamond structure. It is crucial to recognize the pairing motif to classify the crystal phase.

In Fig.~\ref{fig:dimer_diamond} (b), we show the trajectories of particle fractions identified by the classifier trained with or without data augmentation. While the initial and final particle fractions are now the same, the classifier trained with or without data augmentation exhibits different transition behaviors: the non-augmented classifier identifies half of the crystal particles much later than the augmented classifier.

\subsection{Test cases 4 and 5: self-assembly of prolate hexagonal bipyramids}
\label{sec:hexbps}
In the previous section, we showed two different patchy particle systems for which distinguishing the different local environments during assembly is crucial for observing that, in both 2D and 3D, these patchy shapes follow similar assembly pathways. In this section, we focus on two different assembly pathways using geometrically similar building blocks. Both crystals are self-assembled entropically by hard hexagonal bipyramids, but the bipyramid's aspect ratio ($\alpha = 1.28$ vs. $3.0$), defined as the ratio of the particle's height $h$ to its circumcircle diameter of base $d$, i.e., $\alpha=h/d$, greatly influences the intermediate and final products.  For $\alpha=3.0$ orientational order develops prior to translational order, while for $\alpha=1.28$ we observe the opposite.  According to a recent study \cite{lim2023engineering}, the self-assembly pathway from a fluid to the final crystal under slow compression involves an intermediate phase -- a plastic crystal BCT phase for $\alpha=1.28$ and a liquid crystal phase for $\alpha=3.0$.  The final products are BCT and triclinic phases for $\alpha=1.28$ and $\alpha=3.0$, respectively.

\begin{figure*}[!t]
  \centering
  \includegraphics[width=0.9\textwidth]{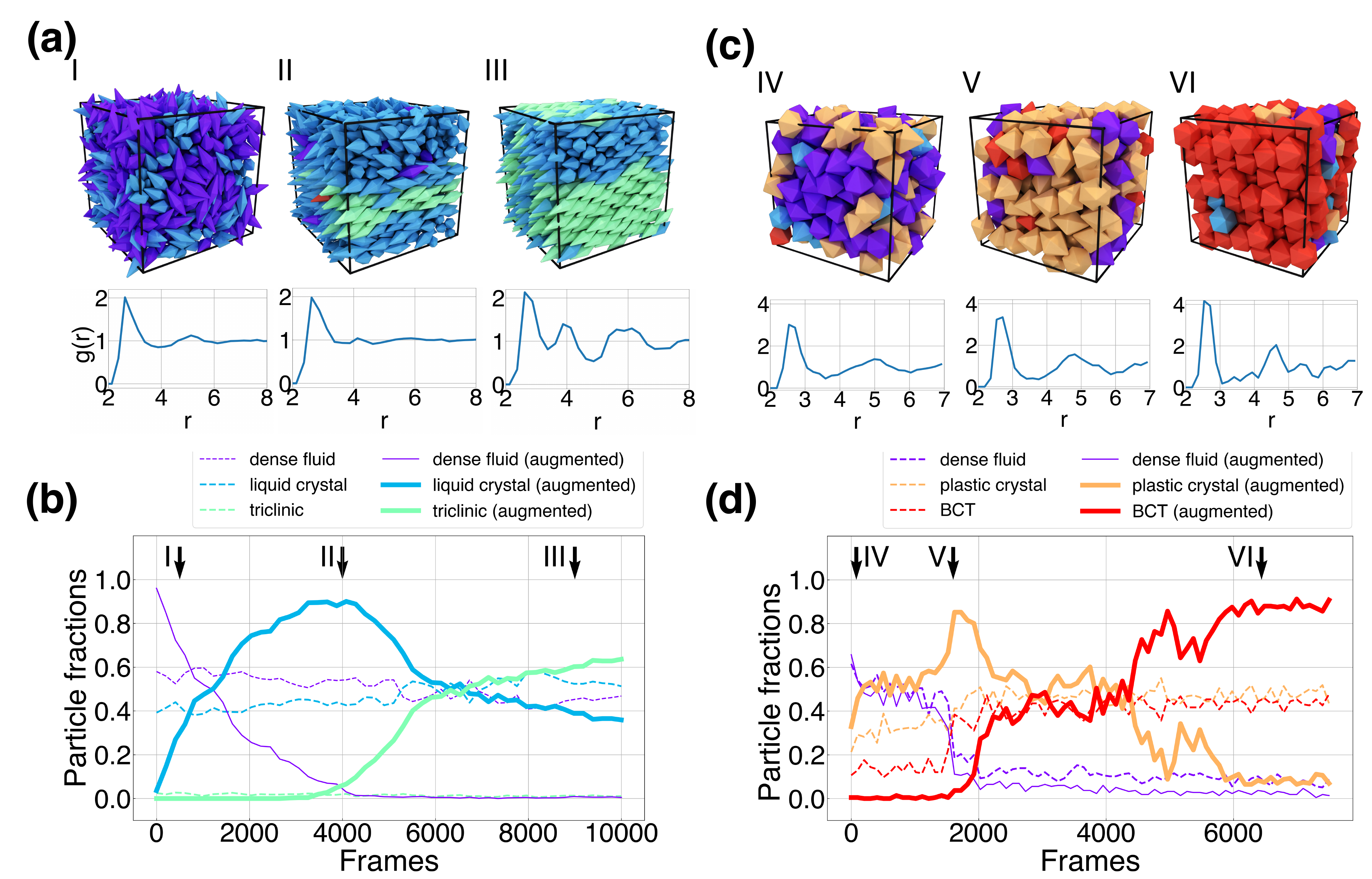}
  \caption
  {\textbf{Hexagonal bipyramid systems} The MLP classifier’s classification results on the assembly trajectories for two systems of hexagonal bipyramids. Both systems exhibit a two-step transition when crystallizing from an initial disordered fluid phase. \textbf{(a)} and \textbf{(c)} The MLP classifier's classification on six snapshots of the hexagonal bipyramid systems with (a) $\alpha = 3.0$ and (b) $\alpha = 1.28$, as well as the corresponding radial distribution functions at the bottom. \textbf{(b)} and \textbf{(d)} The MLP classifier’s classification results on the entire trajectories. The annotations I, II, III, and IV, V, VI correspond to the snapshots in (a) and (b), respectively, for the classifier trained on augmented data. For visualization purposes, solid and dashed lines represent the MLP classifier trained on the data with and without augmentation, respectively.
}\label{fig:hexbps}
\end{figure*}

Because only the aspect ratio, and not symmetry, is different for the two shapes, we prepared only one classifier trained on six different synthetically prepared phases of hexagonal bipyramids for the two aspect ratios. We labeled the disordered phase of both hexagonal bipyramid systems as the same dense fluid phase. Because we consider only the particles' relative positions and quaternions as input, there is no difference in the form of the feature vectors except that they describe different local environments. 

We first show the classification results of the MLP classifier on the test case trajectory of hexagonal bipyramids with $\alpha=3.0$ in Fig. \ref{fig:hexbps} (a) and (b). The dense fluid, liquid crystalline, and triclinic crystal phases are labeled purple, light blue, and light green, respectively. We also calculated the RDF for the three snapshots of Fig. \ref{fig:hexbps} (a). In Fig. \ref{fig:hexbps} (b), our MLP classifier reveals that the system starts to transform into the liquid crystalline phase immediately, followed by slow growth of the triclinic phase. This can also be seen in snapshots I, II, and III of Fig \ref{fig:hexbps} (a). By comparing snapshots I and II, it is evident that the first transition involves only the orientational, and not yet the translational, ordering of particles, leading to a small difference between the two RDFs. Only after the liquid crystalline phase is sufficiently developed does the system order translationally to produce the triclinic crystal; this subsequent behavior is supported by snapshots II and III, as well as their RDFs.

Furthermore, our MLP classifier trained on the augmented data for the hexagonal bipyramids with $\alpha=1.28$ was used to detect a similar two-step transition, and the outcomes are illustrated in Fig. \ref{fig:hexbps} (c) and (d). The plastic crystal and BCT phases are indicated by orange and red colors, respectively. In this instance, the translational ordering of the particles occurs before the orientational ordering, which corresponds to a transition from the dense fluid phase to a plastic crystal phase, followed by a transition to the final BCT phase. It is worth noting that both snapshots IV and V of Fig. \ref{fig:hexbps} (c) possess nearly identical RDFs, meaning the local density and structure are very similar. Despite this, our MLP classifier captures a significant difference in the particle fraction corresponding to dense fluid and plastic crystal. Without data augmentation, the MLP classifier cannot identify particle fractions that match our observed snapshots. In particular, the non-augmented MLP classifier underestimates the particle fractions of the final crystalline phases in both cases. Since the final crystalline phases have both translational and orientational ordering, the MLP classifier performs better with augmented data.

\subsection{Test cases 6 and 7: Self-assembly of truncated polyhedra}
\label{sec:truncated_shapes}
\begin{figure*}[!t]
  \centering
  \includegraphics[width=0.9\textwidth]{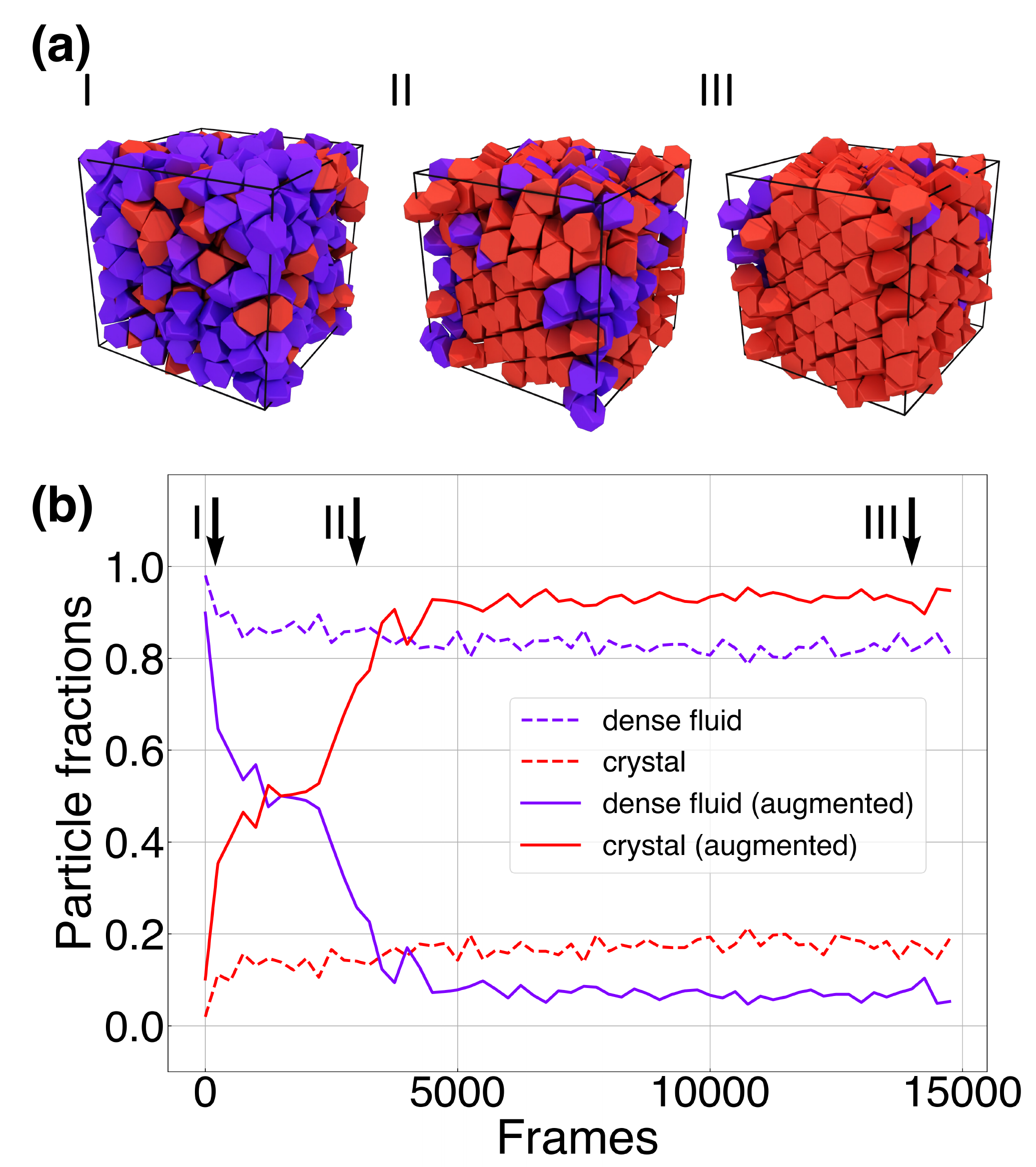}
  \caption
  {\textbf{Diamond structure of truncated tetrahedrons} The MLP classifier's classification results on the self-assembly test trajectory, in which we identify the crystallization of the diamond structure from an initially disordered fluid phase. \textbf{(a)} The MLP classifier’s classification results on snapshots taken at three points along the trajectory. \textbf{(b)} The MLP classifier’s classification results on the entire trajectory. Solid and dashed lines represent the MLP classifier trained on the data with and without augmentation, respectively. The annotations I, II, and III indicate the corresponding snapshots in (a) for the classifier trained on augmented data.
}\label{fig:diamond}
\end{figure*}
\begin{figure*}[!t]
  \centering
  \includegraphics[width=0.9\textwidth]{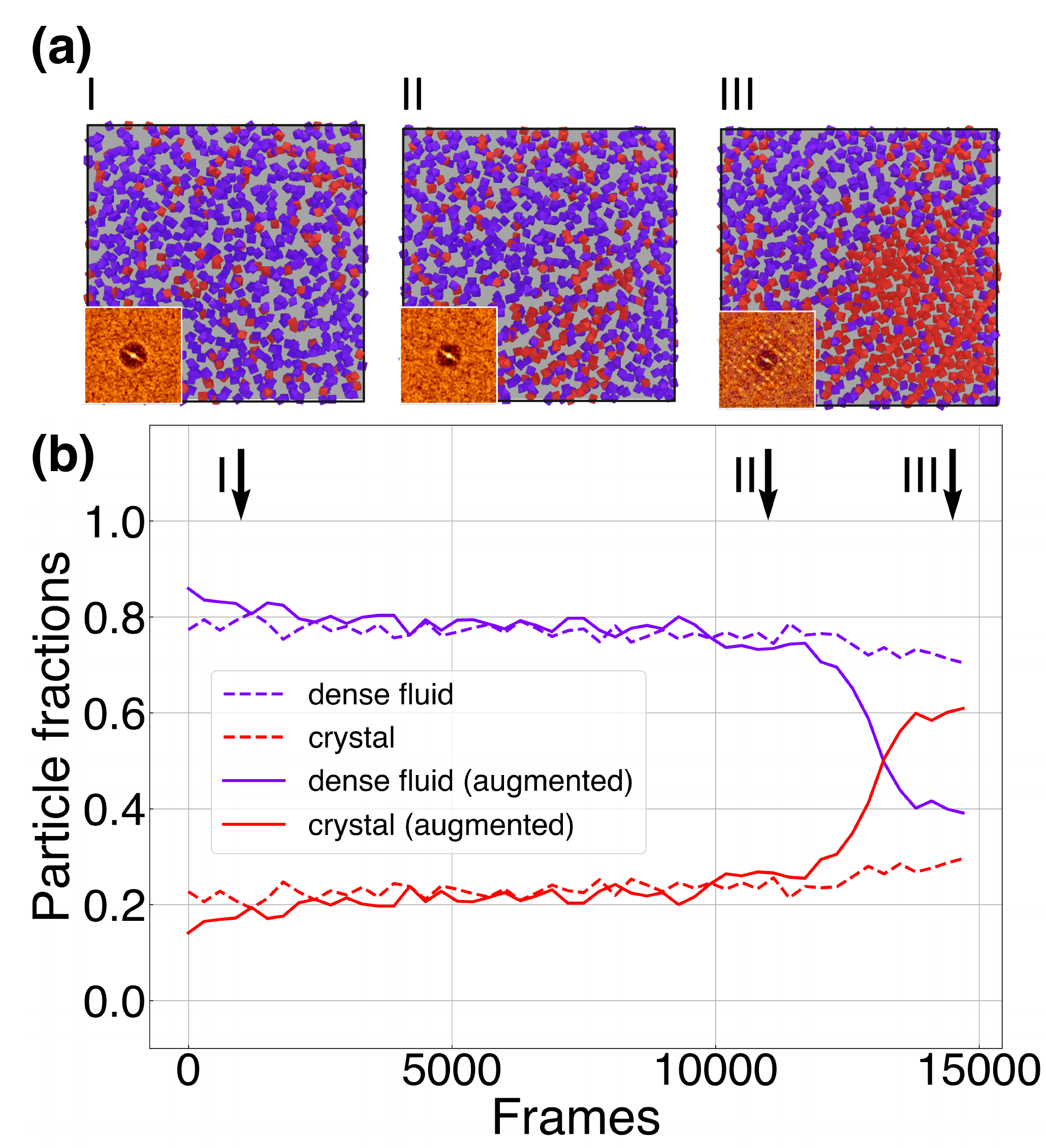}
  \caption
  {\textbf{High-pressure lithium phase of truncated octahedrons} The MLP classifier's classification results on the self-assembly test trajectory, in which we identify the crystallization of the high-pressure lithium phase from an initially disordered fluid phase. \textbf{(a)} The MLP classifier’s classification results on the three snapshots. \textbf{(b)} The MLP classifier’s classification results on the entire trajectory.  Solid and dashed lines represent the MLP classifier trained on the data with and without augmentation, respectively. The annotations I, II, and III indicate the corresponding snapshots in (a) for the classifier trained on augmented data.
}\label{fig:hp_lithium}
\end{figure*}

\noindent As a final test of our classifier, we consider systems that are particularly challenging to identify using order parameters.  As an example, we consider self-assembled systems of truncated tetrahedrons. Damasceno et al.\cite{damasceno2012crystalline} showed that tetrahedrons with varying degrees of truncation self-assemble into a wide range of complex crystal structures, including diamond structures and high-pressure lithium phases. Here, we investigate two different truncated tetrahedron systems with intermediate and large amounts of truncation. Because the tetrahedron gradually transforms to an octahedron with increasing vertex truncation, we refer to the system with intermediate truncation as the truncated tetrahedron and the system with high truncation as the truncated octahedron, to avoid confusion. Our simulations show results consistent with those of Damasceno et al. When we compressed the systems to a packing fraction between $0.5$ and $0.6$, they self-assembled into cubic diamond structures and high-pressure lithium phases for truncated tetrahedrons and truncated octahedrons, respectively.

In Fig.~\ref{fig:diamond} (a) and (b), we demonstrate the classification results of the MLP classifier for the truncated tetrahedrons. In Fig.~\ref{fig:diamond} (b), during the early simulation stage, the classifier trained on augmented data identifies an abrupt increase in the number of particles classified as belonging to a cubic diamond local environment. Subsequently, the diamond structure grows rapidly and remains stable. However, the classifier trained on the data without augmentation reports a slow growth of the diamond structure and fails to identify the phase transition.

In Fig.~\ref{fig:hp_lithium} (b), our MLP classifier trained on augmented data first detected a progressively increasing high-pressure lithium-like local environment before the simulation reached frame12000. After that, our classifier identified that the high-pressure lithium phase reached the critical nucleus size, followed by rapid crystal growth. Thus the MLP classifier discovered a homogeneous nucleation process followed by crystal growth. This nucleation process can also be seen in Fig.~\ref{fig:hp_lithium} (a). From the classification-based coloring in snapshot I, we observed many small sub-critical crystal nuclei form and dissolve in the early stages. In snapshot II, shortly before the rapid growth of the crystal, the MLP classifier identified a noticeable amount of crystal-like local environment in the lower part of the box, which we expect to be the critical nucleus. At the end of the simulation, we can see that the crystal has stabilized in the lower right corner of the box in snapshot III, where it is in coexistence with a dense fluid phase. 

For comparison, we demonstrate the classifier's classification ability without data augmentation. As expected, the MLP classifier trained on the data without augmentation failed to recognize the rapid growth of the crystal. The inability of the classifier to identify the formation of complex crystals is similar to the case of truncated tetrahedrons, as shown in Fig.~\ref{fig:diamond} (a) and (b), and thus for both test cases we see that data augmentation highly improves the performance of the MLP classifier. We can rationalize this performance difference between training on augmented vs. non-augmented data by examining the information contributed by each component of the particle feature vector. Since we started both simulations from very dense fluid phases, there are no significant changes in densities that are encoded in $r_{ij}^{(au)}$ during the formation of crystals. Therefore, the information provided by $r_{ij}^{(au)}$ is insufficient to recognize the formation of crystals. As a result, whether the MLP classifier can correctly learn to identify the formation of crystals largely depends on whether $\theta_{ij}$, $\phi_{ij}$ and $\mathbf{q}_{ij}$ are augmented.

\section*{Conclusions}
\label{sec:conclusion}

To summarize, we have developed a simple, yet powerful, physics-agnostic local environment classifier specifically designed for systems of particle shapes utilizing a multilayer perceptron. As demonstrated in this paper, our method is applicable to a range of enthalpically and entropically patchy particle systems. Importantly, our MLP classifier does not need conventional roto-translation invariant symmetry functions to transform per-particle quantities to input descriptors. Instead, it directly takes particle positions and orientations as input features, complemented by shape-encoded data augmentation. To demonstrate robustness and flexibility, our classifier's performance was assessed on a variety of self-assembling systems, including hard cubes, 2D and 3D patchy shapes, hexagonal bipyramids with two different aspect ratios, and two different truncated shapes. The data augmentation method we used is straightforward and easily transferable to other systems involving particle orientations, such as molecular or coarse-grained systems. As a result, our approach should be useful in classifying different polymorphs formed by molecules by properly defining orientation through quaternions. Furthermore, due to the simplicity and the promising classification performance, our method may be applied to study nucleation pathways of colloidal systems with biased simulation techniques, such as umbrella sampling and metadynamics\cite{dietrich2023machine}. Such applications will be explored in future works.

\section{Conflicts of interest}
\label{sec:conflicts_of_interest}
There are no conflicts to declare.

\section{Acknowledgements}
\label{sec:acknowledgements}

This research was supported by a CDS\&E grant from the National Science Foundation (NSF), Division of Materials Research Award No. DMR 2302470. This work used resources from the Extreme Science and Engineering Discovery Environment (XSEDE), supported by the same aforementioned NSF grant. Computational resources and services were supported by Advanced Research Computing at the University of Michigan, Ann Arbor. The authors would like to thank Yuan Zhou for providing data on patchy particle systems and for helpful discussion on the symmetry of shapes; Michael Engel for providing the crystal data for truncated octahedrons; Thi Vo's help in setting up the MD simulations; and Brandon Butler, Ziyue Zou, and Yihang Wang for their helpful suggestions.

\section*{Data Availability}
\label{sec:dataavailability}

The data that support the findings of this study are available from the corresponding authors upon reasonable request.

\bibliography{references.bib}

\end{document}